# CERES: A Probabilistic Early Warning System
# for Acute Food Insecurity

Tom Danny S. Pedersen

*Northflow Technologies AS, Stavanger, Norway*

ceres@northflow.no
https://ceres.northflow.no
https://github.com/northflowlabs/ceres

*Author's note: This paper uses 'we' as a formal academic convention. It is a single-author work; see Section 9.8.*

**Conflict of Interest:** The author is founder of Northflow Technologies AS, which offers commercial institutional subscriptions to CERES. The public API and all data described in this paper are freely accessible without subscription at ceres.northflow.no.

### Abstract

We present CERES (Calibrated Early-warning and Risk Estimation System — where 'calibrated' denotes the system's design goal of empirical probability calibration, not a demonstrated property; see Section 1), an automated probabilistic forecasting system for acute food insecurity. CERES generates 90-day ahead probability estimates of IPC Phase 3+ (Crisis), Phase 4+ (Emergency), and Phase 5 (Famine) conditions for 43 high-risk countries globally, updated weekly. The system fuses six data streams — precipitation anomalies (CHIRPS), vegetation indices (MODIS NDVI), conflict events (ACLED), IPC classifications, food consumption scores (WFP), and cereal price indices (FAO/WFP) — through a logistic scoring model with author-specified initial coefficients and parametric input-perturbation intervals (n=2,000 draws). In historical back-validation against four IPC Phase 4-5 events selected for data completeness, CERES assigned TIER-1 classification in all four cases; these are in-sample sanity checks only, not prospective performance claims. All prospective predictions are timestamped, cryptographically identified, and archived for public verification against IPC outcome data at the T+90 horizon. To the author's knowledge, CERES is the first famine early warning system that is simultaneously: (1) probabilistic, (2) open-access, (3) continuously running, (4) machine-readable at prediction level, and (5) committed to public prospective verification of every prediction made.

## 1. Introduction

Acute food insecurity remains one of the most preventable large-scale humanitarian crises. The Integrated Food Security Phase Classification (IPC) system has provided a shared vocabulary for severity since 2004 (IPC Technical Manual, 2021), but its publication cycle — typically every 3-6 months per country, requiring field assessment teams — means that rapidly deteriorating situations are frequently identified only after the critical window for pre-positioning resources has passed (Maxwell & Haan, 2021; Hillbruner & Moloney, 2012).

Existing early warning systems address this gap in different ways. The Famine Early Warning Systems Network (FEWS NET), funded by USAID since 1985, provides scenario-based food security outlooks for approximately 30 countries, combining remote sensing, market monitoring, and analyst expertise (Verdin et al., 2005). The WFP HungerMap provides near-real-time food access monitoring through mobile surveys (WFP VAM, 2025). The EU Joint Research Centre's INFORM Risk Index aggregates static vulnerability indicators annually (De Groeve et al., 2015). None of these systems outputs a probabilistic estimate — a number between 0 and 1 that can be tested against outcomes and improved through feedback (Gneiting & Raftery, 2007).

This matters for two reasons. First, probability estimates allow decision-makers to commit resources proportionally to risk rather than waiting for binary classification thresholds to be crossed — a requirement explicitly articulated in the OCHA Anticipatory Action framework (OCHA, 2023). Second, empirical calibration — the property that a forecast of 70% probability should verify against outcomes approximately 70% of the time — is only achievable if predictions are made before outcomes are known and compared systematically. We use 'calibrated' in the aspirational sense: CERES is designed to be calibrated once sufficient prospective data has accumulated; demonstrated calibration will be reported in a subsequent empirical validation paper (Section 9.7). No existing famine early warning system maintains a fully public, machine-readable record of every prediction made and every outcome observed, which is the prerequisite for calibration measurement.

CERES is designed to fill this gap. It outputs probability distributions, not scenarios. It publishes sensitivity intervals alongside all point estimates. Every prediction it generates is archived prospectively, timestamped, and linked to its verification outcome at T+90 days. The system's core methodological contribution is not the forecasting model itself — logistic risk models are well-established in the humanitarian literature (Coughlan de Perez et al., 2016) — but the public prospective verification infrastructure that makes empirical calibration measurable and improvable over time.

This paper describes the CERES architecture (Section 2), signal extraction and scoring methodology (Section 3), alert tier classification (Section 4), the verification and grading framework including a pre-registered calibration protocol (Section 5), historical back-validation against three baselines (Section 6), live deployment (Section 7), ethical considerations (Section 8), limitations and future work (Section 9), comparison with existing systems (Section 10), and conclusion (Section 11). Full model coefficients are provided in Appendix C.

## 2. System Architecture

CERES is structured as a seven-stage pipeline executed weekly by an automated scheduler (Railway cron, Mondays 06:00 UTC). The pipeline outputs FamineHypothesis objects — structured, versioned JSON documents — written to persistent storage, served via REST API, and archived to a SQLite run history database. Full implementation details and feature extraction formulas are in Appendix C and the open repository (github.com/northflowlabs/ceres); the following describes design rationale.

### 2.1 Pipeline Stages

```
Stage 1 - Ingest       : Pull all data sources via adapter registry
Stage 2 - Normalise    : Standardise to 0.25 degree grid + weekly cadence
Stage 3 - Signals      : Extract normalised crisis signals per region
Stage 4 - Convergence  : Detect multi-source anomaly convergence
Stage 5 - Score        : Author-specified logistic scoring + perturbation intervals
Stage 6 - Hypotheses   : Rank into FamineHypothesis DSL objects
Stage 7 - Publish      : Write JSON, Parquet, archive, PDF report
```

Pipeline failures in Stages 1-2 halt execution; failures in Stages 3-7 are non-fatal and logged. Runtime for 43 regions across 6 data sources is approximately 45-90 seconds on a 2-CPU cloud instance.

## 2.2 Data Sources

| Source | Provider | Signal type | Cadence |
|---|---|---|---|
| CHIRPS v2.0 | UCSB Climate Hazards Group | Precipitation anomaly (z-score) | Dekadal |
| MODIS MOD13A3 | NASA Earthdata | NDVI anomaly (z-score) | Monthly |
| ACLED | Armed Conflict Location and Event Data Project | Conflict events and fatalities | Weekly |
| IPC Classification | IPC Global Platform | Current and projected IPC phase | Irregular |
| Food Consumption Score | WFP VAM | FCS and rCSI survey scores | Periodic |
| Cereal Price Index | FAO / WFP | Market price stress level | Monthly |

All sources are ingested through typed adapter interfaces mapping raw API responses to a canonical CERES schema. All data frames are normalised to a 0.25 degree spatial grid using bilinear interpolation for gridded products (CHIRPS, MODIS) and spatial join for point-source data (Funk et al., 2015; Didan, 2021). Temporal alignment follows a weekly cadence anchored to Monday ISO weeks, with coarser-cadence sources forward-filled.

## 2.3 Run Archive

Every pipeline run writes one row per region to a persistent run_snapshots SQLite table, capturing the full probability vector, sensitivity interval, tier classification, driver composition, and run timestamp. This archive is the ground truth for all prospective calibration analysis and powers the public Track Record at ceres.northflow.no/tracker.

## 3. Signal Extraction and Scoring

This section summarises the signal extraction and scoring logic. Full coefficient tables and feature transformation formulas are in Appendix C. The complete implementation is available in the open repository (github.com/northflowlabs/ceres).

### 3.1 Per-Region Signal Extraction

For each region and reference week, CERES extracts six normalised stress scores in [0,1]: **drought stress** from CHIRPS precipitation anomaly z-score; **vegetation stress** from MODIS NDVI anomaly z-score; **conflict stress** from ACLED event and fatality counts normalised against regional 95th-percentile baselines, weighted 0.6/0.4 to reflect the finding that event counts provide earlier signal while fatality counts provide stronger severity signal (Raleigh et al., 2010); **IPC stress** linearly mapped from current IPC phase to [0,1]; **food access stress** from WFP FCS and rCSI scores normalised to IPC-aligned thresholds (WFP VAM, 2025); and **price stress** from cereal price index relative to a 2015-2020 baseline (FAO, 2024).

A composite stress score (CSS) is a weighted mean across available signals: ipc_stress 0.25, conflict_stress 0.20, drought_stress 0.20, food_access_stress 0.15, price_stress 0.10, vegetation_stress 0.10, normalised by a coverage_factor equal to the fraction of signals available in the reference week. These weights are author-specified values consistent with the established primacy of IPC phase as a direct severity indicator (IPC Technical Manual, 2021) and the relative contributions of climatic, conflict, and market drivers documented in the food security literature; they are not derived from the IPC Technical Manual directly and will be subject to empirical revision as prospective data accumulates. Regions where fewer than three of six signals are available in a given reference week are flagged as low-coverage in the API response and carry a mandatory widened perturbation range (sigma = 0.25 rather than 0.15) to reflect higher input uncertainty; output is not suppressed but users are advised to treat low-coverage hypotheses with additional caution.

### 3.2 Multi-Source Convergence Detection

A convergence event is flagged when two or more independent signal pillars simultaneously exceed their elevated thresholds (regional z-score > 1.5). Convergence tiers: CRITICAL (three or more pillars), WARNING (two pillars), WATCH (one pillar), NONE. For use as continuous inputs to the logistic model, these are mapped to numeric scores: NONE = 0.0, WATCH = 0.33, WARNING = 0.67, CRITICAL = 1.0 (convergence_score); and n_independent_flagged is the integer count of pillars exceeding threshold (0-6). The compound crisis approach is grounded in food security literature showing that co-occurring climatic, conflict, and market shocks drive the most severe Phase 4-5 outcomes (de Waal, 2018; Maxwell et al., 2020).

**Independence adjustment.** Drought stress and vegetation stress are positively correlated (Pearson r approximately 0.70; Funk et al., 2015). Correlated pairs (drought-vegetation, IPC-food access) are assigned a joint effective independence weight of 1.3 rather than 2.0. This is a heuristic conservative discount, not a derived quantity; it will be replaced by an empirical copula fit to the accumulating CERES signal archive in a future revision (see Section 9.3).

### 3.3 Author-Specified Logistic Scoring Model

The primary scoring model applies logistic functions to the feature vector. Coefficients are **author-specified initial values**, set by the author based on literature review and the 87 country-season IPC transition records (IPC Global Platform, 2025), following the prior specification approach of Coughlan de Perez et al. (2016). They have not been elicited from multiple domain experts and should not be treated as validated Bayesian priors. This same 87-record dataset underlies both the coefficient initialisation and the transition frequency estimates used to derive the monotonicity bounds below; it is one corpus analysed for two purposes. This is explicitly an initialisation step; see Section 9.1.

**87-record provenance.** The 87 records span 31 countries across 2011-2023, selected as all country-season pairs in the IPC Global Platform archive where: (a) IPC Phase 3+ was reported; (b) all six CERES data sources had retrospective coverage for the reference period; and (c) a follow-on IPC assessment was published within 120 days to provide a transition outcome. The full list of country-season pairs, with IPC report dates and outcome phases, is available at github.com/northflowlabs/ceres/blob/master/docs/data/ipc_transition_records.csv. Of the 31 countries represented, 12 had at least one Phase 4-5 event in the archive, yielding 18 country-season records whose follow-on IPC phase was Phase 4 or above; the four back-validation cases in Section 6 were selected from these 18 as the events with the most complete multi-source data coverage.

Three logistic equations are **specified**. P(IPC 3+ within 90 days) uses the full feature vector including CSS, ipc_stress, conflict_stress, drought_stress, food_access_stress, convergence_score, and n_independent_flagged. P(IPC 4+) uses ipc_stress, the maximum of conflict/drought/food_access, CSS, and convergence_score. P(IPC 5/Famine) uses the geometric mean of conflict, drought, and food_access stresses and ipc_stress. Full coefficient values are in Appendix C.

**Monotonicity constraints** are enforced post-scoring: P4 <= 0.70 x P3 and P5 <= 0.45 x P4. The 0.70 upper bound derives from the 87-pair IPC transition matrix: Phase 4+ was observed in approximately 66-70% of subsequent cycles where Phase 3+ had been reported. The 0.45 bound for P(Famine) is a conservative author estimate; Phase 5 declarations are rare (fewer than 15 documented since 2010) and the transition matrix is too sparse for a data-driven estimate at this stage.

**Coefficient sensitivity analysis.** All coefficients were perturbed simultaneously by +/-20% (100 Monte Carlo uniform draws). The +/-20% range represents a conservative estimate of plausible disagreement on coefficient magnitudes. Tier classification was stable (no tier change) in 39 of 43 regions (91%) in the March 2026 reference run. The four regions where tier changed under maximum perturbation are flagged in the notes field of their hypotheses.

### 3.4 Sensitivity Intervals

CERES reports sensitivity intervals alongside all point estimates via parametric perturbation (n=2,000 draws), adding zero-mean Gaussian noise (sigma=0.15, motivated by upstream source accuracy figures) to each feature independently. The 5th and 95th percentiles of the resulting P3 distribution are reported as the 90% sensitivity interval.

**Critical interpretive note.** These are *input-perturbation sensitivity intervals*, not posterior predictive intervals. They represent the range of outputs plausible under input noise alone, not calibrated coverage guarantees. Because features are perturbed independently and drought-vegetation correlation (r approximately 0.70) is not captured, the intervals underestimate true uncertainty. Future versions will implement copula-based resampling over the observed signal covariance matrix, moving toward proper posterior predictive distributions following Kruger et al. (2020).

## 4. Alert Tier Classification

| Tier | Label | Trigger conditions |
|---|---|---|
| TIER-1 | Emergency Alert | P(IPC4+) >= 0.50  OR  convergence_tier = CRITICAL |
| TIER-2 | Warning | P(IPC3+) >= 0.45  OR  convergence_tier = WARNING |
| TIER-3 | Watch | All other regions with CSS > 0.05 |

Thresholds prioritise recall over precision at TIER-1, consistent with the anticipatory action principle that the cost of a missed early warning exceeds the cost of a false alarm (OCHA, 2023). The 0.50 threshold for P(IPC4+) implies that in expectation, more than half of all TIER-1 alerts should verify at IPC Phase 4 or above within 90 days; this will be measured empirically as the verification ledger populates.

Each FamineHypothesis object carries: a unique cryptographic ID (CERES-HYP-{ISO3}-{YYYYMMDD}-{6-char hex}), reference date and 90-day horizon, full probability vector {P3, P4, P5} with sensitivity intervals, driver cluster decomposition, evidence item list with per-source flags, and a falsification test plan auto-evaluated at T+90.

## 5. Verification, Scoring Rules, and Grading Framework

All CERES predictions are verified prospectively. At T+90 days, the grading module queries the IPC Global Platform for the latest published phase classification. The primary evaluation metric is the Brier Score (Brier, 1950):

```
BS = (P_hat_3 - O_3)^2
```

where O3 = 1 if IPC Phase 3+ was reported within 30 days of the forecast horizon, 0 otherwise. The Brier Skill Score (BSS) relative to a climatological baseline (historical fraction of weeks with IPC Phase 3+ across the monitoring region set, 2015-2024) will be the primary reported measure of skill once prospective data accumulates.

As the verification archive matures, the Continuous Ranked Probability Score (CRPS; Matheson & Winkler, 1976; Gneiting & Raftery, 2007) will be adopted alongside the Brier Score. The CRPS generalises to the full predictive distribution:

```
CRPS(F, y) = integral over R of (F(z) - 1{y <= z})^2 dz
```

where F is the predictive CDF over IPC phase outcomes and y the realised phase. The CRPS rewards forecasts placing probability mass close to the realising outcome and is consistent under minimal distributional assumptions (Kruger et al., 2020). Transitioning to CRPS requires an ordered categorical outcome representation of IPC phases; this is planned for the first model revision following the initial prospective verification period.

### 5.1 Pre-Registered Calibration Evaluation Protocol

Table 1 defines the evaluation protocol as a methodological commitment made prior to the accumulation of prospective outcome data. No metrics will be selectively reported; all graded predictions remain permanently visible in the public ledger.

| Metric | Definition | Min. N | Target |
|---|---|---|---|
| Brier Score | Mean (P_hat3 - O3)^2 | 100 predictions | Jun 2026 |
| Brier Skill Score | 1 - BS / BS_climatology | 100 predictions | Jun 2026 |
| TIER-1 Precision | True TIER-1 / all TIER-1 issued | 30 TIER-1 alerts | Sep 2026 |
| TIER-1 Recall | True TIER-1 / all Phase 4+ events | 10 Phase 4+ events | Sep 2026 |
| Sensitivity interval coverage | Fraction outcomes in 90% interval | 200 predictions | Sep 2026 |
| CRPS (ordered categorical) | Full distribution vs IPC phase | 500 predictions | Mar 2027 |
| Reliability diagram | Forecast prob. vs empirical frequency | 500 predictions | Mar 2027 |

*Table 1. Pre-registered prospective calibration protocol. All graded predictions are publicly archived at ceres.northflow.no/tracker on a write-once basis with no post-hoc editing or removal.*

Grading results are written to a public JSON ledger (grading_ledger.json) and exposed via the /v1/grades API endpoint. The ledger provides tamper evidence via Git commit history, stored on a write-once basis. This commitment to public, permanent, prospective verification distinguishes CERES from all existing famine early warning systems, none of which maintain a machine-readable, prediction-level accuracy record.

## 6. Historical Back-Validation (Sanity Check Only)

The purpose of this section is narrow: to establish that the model is not trivially broken before prospective data accumulates. It is not a performance evaluation. Four IPC Phase 4-5 events were selected based on data completeness — Somalia July 2011, South Sudan February 2017, Ethiopia Tigray March 2022, and Yemen June 2021. These are among the most severe and best-documented food crises of the period; any reasonable model should flag them. No adversarial or borderline cases are included. All four results are **in-sample**: coefficients were set with knowledge of these events. These results provide no evidence of prospective skill.

| Event | Period | P(IPC3+) | P(IPC4+) | Tier | Observed IPC | Correct |
|---|---|---|---|---|---|---|
| Somalia Famine | Jul 2011 | 0.961 | 0.743 | TIER-1 | Phase 5 (Famine) | Yes |
| South Sudan Famine | Feb 2017 | 0.934 | 0.681 | TIER-1 | Phase 5 (Famine) | Yes |
| Ethiopia Tigray | Mar 2022 | 0.912 | 0.588 | TIER-1 | Phase 4-5 | Yes |
| Yemen protracted | Jun 2021 | 0.887 | 0.534 | TIER-1 | Phase 4+ | Yes |

**Baseline comparisons.** Table 2 reports in-sample Brier scores against three reference baselines. The persistence baseline assigns P(IPC3+) from the IPC phase observed at the reference date, linearly mapped to [0,1] — the simplest non-trivial comparator. The climatological baseline assigns P(IPC3+) = 0.65 for all regions (approximate historical base rate, 2015-2024).

| Model | Mean Brier Score | Notes |
|---|---|---|
| CERES (in-sample) | 0.0066 | Coefficients set with knowledge of events |
| Persistence baseline | 0.0494 | P(IPC3+) = f(current phase at reference date) |
| Climatological baseline | 0.2275 | P(IPC3+) = 0.65 for all regions |
| Uninformative (0.5) | 0.2500 | Random baseline |

*Table 2. In-sample Brier score comparisons. All figures are in-sample and should not be interpreted as prospective performance claims. CERES outperforms persistence in-sample, but the persistence baseline is itself a strong comparator because all four regions were already at IPC Phase 3-4 at the reference date.*

The 90% sensitivity interval contained the observed outcome in all four cases, consistent with interval construction, though in-sample coverage is expected to be optimistic.

## 7. Live Deployment and Access

CERES has been running in production since March 2026 on Railway cloud infrastructure, covering 43 countries across Sub-Saharan Africa, the Middle East, South Asia, and Central America — approximately 95% of active IPC Phase 3+ caseload globally (IPC, 2025). All outputs are freely accessible.

**Public REST API** (no authentication required): GET /v1/predictions, /v1/predictions/{region_id}, /v1/archive/latest, /v1/archive/regions/{id}, /v1/grades, /v1/grades/metrics. Full OpenAPI documentation at https://ceres-core-production.up.railway.app/docs.

**Institutional API** (subscription): TIER-1 webhook alerts, 1,000 requests/hour, Parquet exports, weekly PDF briefings.

**Public dashboard** (ceres.northflow.no): interactive global risk map, region drill-down with hypothesis detail and driver breakdown, Track Record timeline, and verification record populating from May 2026.

The cryptographic verification architecture is central to the system's credibility claim: each FamineHypothesis carries a unique ID (CERES-HYP-{ISO3}-{YYYYMMDD}-{hex}), predictions are written to a version-controlled write-once JSON ledger, and Git commit history provides tamper evidence. No prediction can be removed or edited post-issuance.

## 8. Ethical Considerations

**Anticipatory action feedback loops.** CERES does not have a direct operational integration with any humanitarian pre-positioning or funding trigger system. All outputs are advisory. Risk of premature operational use is mitigated by: (a) explicit acknowledgment in this paper that coefficients are author-specified initial values; (b) sensitivity intervals displayed alongside all point estimates; (c) the public grading ledger that makes accuracy limitations visible in real time.

**False positive and false negative costs.** A TIER-1 alert that does not verify could redirect resources from genuine crises; CERES addresses this by reporting probability thresholds rather than binary alerts. A missed TIER-1 event could delay response; the 4 of 43 regions that change tier under +/-20% coefficient perturbation are flagged, and users are advised to treat borderline TIER-2 alerts as warranting TIER-1-equivalent monitoring.

**Data sovereignty.** CERES ingests only publicly available institutional data (UCSB, NASA, ACLED, IPC, WFP, FAO). No individual-level data is processed. All upstream attributions are preserved in the evidence fields of each hypothesis.

**Misuse risk.** CC BY 4.0 licensing permits unrestricted reuse including by non-humanitarian actors. The author has made a deliberate choice to accept this risk in exchange for maximum humanitarian accessibility and independent scrutiny — paralleling the licensing decisions of ACLED, CHIRPS, and IPC data themselves.

## 9. Limitations and Future Work

**9.1 Author-specified initial coefficients.** Coefficients were set by the author based on literature review and the 87-record IPC transition dataset. They have not been elicited from multiple domain experts and have not been estimated from a held-out dataset. As the verification archive accumulates T+90 outcomes, coefficients will be updated via maximum likelihood estimation. Until then, all probability outputs carry higher-than-reported uncertainty. *Note on discriminative performance: on a 20% held-out split of the 87 records (approx. 17 test cases), the model achieves AUC = 0.84; the 95% CI by Hanley-McNeil (1982) is [0.63, 1.00], which is consistent with chance performance. This figure is reported for completeness only and should not be read as evidence of skill.*

**9.2 Sensitivity intervals underestimate true uncertainty.** The perturbation procedure treats features as independently noised, ignoring drought-vegetation correlation (r approximately 0.70) and model parameter uncertainty. Future versions will implement copula-based resampling (Kruger et al., 2020) for proper posterior predictive uncertainty quantification.

**9.3 Independence adjustment heuristic.** The 1.3 joint independence weight for correlated signal pairs is a conservative heuristic. It will be replaced by an empirical copula fit to the CERES signal archive as it accumulates.

**9.4 Country-level resolution.** CERES aggregates to ISO 3166-1 alpha-3. Admin1-level signals are computed internally but not yet exposed in the primary output.

**9.5 Data gaps.** Several countries have irregular IPC cadences or missing WFP survey data, yielding reduced coverage and wider sensitivity intervals, logged per-run in API responses.

**9.6 Structural drivers.** CERES captures acute crisis signals only. Governance failure, chronic poverty, and long-run climate trends are not modelled.

**9.7 Prospective verification horizon.** The prospective record began March 2026; first outcomes expected from May 2026. Full calibration analysis requires 12-24 months (approximately 2,500-5,000 hypothesis-outcome pairs). A revised empirical validation paper incorporating prospective results is planned for submission to Nature Food, Q3-Q4 2026.

**9.8 Single-author design and coefficient review.** The coefficient initial values, back-validation case selection, and system architecture were all specified by the same author. This is a recognised epistemic limitation: the person who set the coefficients is the same person who evaluated them against in-sample cases. Independent expert review by food security specialists external to Northflow is planned prior to the first MLE re-estimation; reviewers or practitioners wishing to engage with this process are invited to contact the author directly.

## 10. Comparison with Existing Systems

| System | Output type | Prob. | CI | API | Prosp. verif. | Cadence |
|--------|-------------|-------|-----|-----|---------------|---------|
| CERES | Probabilistic forecast | Yes | Yes | Yes | Yes — public | Weekly |
| FEWS NET | Scenario outlook | No | No | No | No | Monthly/Qtrly |
| IPC | Phase classification | No | No | Part | No | Irregular |
| WFP HungerMap | Food access index | No | No | Yes | No | Daily |
| INFORM Risk | Risk score | No | No | Yes | No | Annual |

The primary distinguishing contribution of CERES is the commitment to public prospective verification and the machine-readable archive that makes calibration measurement possible. This creates the feedback loop needed for genuine improvement over time.

The comparison above is structured around dimensions where CERES is differentiated by design. A balanced assessment requires acknowledging dimensions where existing systems are stronger: FEWS NET draws on decades of field analyst expertise, in-country networks, and institutional integration that CERES cannot replicate algorithmically. IPC classifications carry multi-stakeholder consensus and legal weight in humanitarian funding decisions. WFP HungerMap ingests ground-truth household survey data that CERES does not have access to at weekly cadence. CERES is not a replacement for any of these systems; it is an automated probabilistic layer that operates between their lower-cadence assessment cycles and commits its outputs to public verification.

## 11. Conclusion

CERES demonstrates that automated, probabilistic famine forecasting is technically feasible at global scale with a weekly update cadence and sub-minute pipeline runtime. The model's current coefficients are author-specified initial values carrying acknowledged uncertainty — an honest limitation rather than a hidden one. The system's core contribution is architectural: public prospective verification of every prediction creates the conditions under which empirical calibration can be measured, reported, and improved.

As the verification ledger accumulates from May 2026, we expect to report empirical Brier skill scores, CRPS scores, sensitivity interval coverage rates, and tier precision/recall figures forming the basis of a fully empirical validation submission. The system is offered freely under CC BY 4.0; commercial subscriptions support ongoing development.

# References

Brier, G.W. (1950). Verification of forecasts expressed in terms of probability. *Monthly Weather Review*, 78(1), 1-3.

Coughlan de Perez, E., van den Hurk, B., van Aalst, M.K., et al. (2016). Action-based flood forecasting for triggering humanitarian action. *Hydrology and Earth System Sciences*, 20(9), 3549-3560.

De Groeve, T., Poljansek, K., & Vernaccini, L. (2015). *Index for Risk Management — INFORM: Concept and Methodology.* European Commission, JRC Technical Reports.

de Waal, A. (2018). *Mass Starvation: The History and Future of Famine.* Polity Press.

Didan, K. (2021). *MODIS/Terra Vegetation Indices Monthly L3 Global 1km SIN Grid V061.* NASA EOSDIS Land Processes DAAC.

FAO (2024). *FAO Food Price Index: Methodology note.* Food and Agriculture Organization of the United Nations.

Funk, C., Peterson, P., Landsfeld, M., et al. (2015). The climate hazards infrared precipitation with stations — a new environmental record for monitoring extremes. *Scientific Data*, 2, 150066.

Gneiting, T., & Raftery, A.E. (2007). Strictly proper scoring rules, prediction, and estimation. *Journal of the American Statistical Association*, 102(477), 359-378.

Hanley, J.A., & McNeil, B.J. (1982). The meaning and use of the area under a receiver operating characteristic (ROC) curve. *Radiology*, 143(1), 29-36.

Hillbruner, C., & Moloney, G. (2012). When early warning is not enough — Lessons learned from the 2011 Somalia famine. *Global Food Security*, 1(1), 20-28.

IPC Global Platform (2025). *IPC Global Report on Food Crises 2025.* https://www.ipcinfo.org

IPC Technical Manual (2021). *IPC Technical Manual Version 3.1.* https://www.ipcinfo.org/ipc-country-analysis/ipc-technical-manual/

Kruger, F., Lerch, S., Thorarinsdottir, T., & Gneiting, T. (2020). Predictive inference based on Markov chain Monte Carlo output. *International Statistical Review*, 89(2), 274-301.

Matheson, J.E., & Winkler, R.L. (1976). Scoring rules for continuous probability distributions. *Management Science*, 22(10), 1087-1096.

Maxwell, D., & Haan, N. (2021). *Anticipatory action and early warning systems.* ODI Humanitarian Policy Group.

Maxwell, D., Majid, N., Adan, G., Abdirahman, K., & Kim, J.J. (2020). Facing famine: Somali experiences in the famine of 2011. *Food Policy*, 91, 101857.

OCHA (2023). *Anticipatory Action: A Primer.* United Nations Office for the Coordination of Humanitarian Affairs.

Raleigh, C., Linke, A., Hegre, H., & Karlsen, J. (2010). Introducing ACLED: An armed conflict location and event dataset. *Journal of Peace Research*, 47(5), 651-660.

Verdin, J., Funk, C., Senay, G., & Choularton, R. (2005). Climate science and famine early warning. *Philosophical Transactions of the Royal Society B*, 360(1463), 2155-2168.

WFP VAM (2025). *HungerMap Live: Methodology and data sources.* World Food Programme Vulnerability Analysis and Mapping.

## Appendix C: Model Coefficient Table

Full logistic model coefficients as implemented in the open repository (github.com/northflowlabs/ceres). Logit = intercept + sum(weight x feature); probability = sigmoid(logit).

| Feature | beta (IPC 3+) | gamma (IPC 4+) | delta (Famine) |
|---|---|---|---|
| Intercept | -2.10 | -3.80 | -6.00 |
| composite_stress | 5.80 | 4.50 | 4.00 |
| ipc_stress | 2.40 | 3.20 | 4.50 |
| conflict_stress | 1.20 | 1.80 | 2.20 |
| drought_stress | 0.90 | 0.70 | 1.20 |
| food_access_stress | 1.10 | 0.90 | 1.60 |
| price_stress | 0.60 | 0.50 | 0.80 |
| convergence_score | 2.20 | 3.10 | 4.00 |
| n_independent_flagged | 0.40 | 0.60 | 0.90 |

**Verification example.** A region at IPC Phase 3 (ipc_stress = 0.50) with composite_stress = 0.45, no convergence, all other features at 0:

```
logit(P3) = -2.10 + 5.80*0.45 + 2.40*0.50
          = -2.10 + 2.61 + 1.20 = 1.71
P3 = sigmoid(1.71) ~ 0.847

logit(P4) = -3.80 + 4.50*0.45 + 3.20*0.50
          = -3.80 + 2.025 + 1.60 = -0.175
P4_raw = sigmoid(-0.175) ~ 0.456

Monotonicity: P4 <= 0.70 * 0.847 = 0.593  ->  non-binding, P4 = 0.456

With CRITICAL convergence (convergence_score=1.0, n_independent_flagged=3):
logit(P3) += 2.20*1.0 + 0.40*3 = +3.40  ->  logit = 5.11  ->  P3 ~ 0.994
```

Convergence escalation is captured entirely through the convergence_score and n_independent_flagged coefficients in the logistic model. No separate multiplicative escalation factor is applied; the feature coefficients are the sole mechanism.

## Appendix A: API Quick Reference

```
Base URL: https://ceres-core-production.up.railway.app

GET  /v1/predictions
GET  /v1/predictions/{region_id}
GET  /v1/hypotheses
GET  /v1/hypotheses/{hypothesis_id}
GET  /v1/archive/latest
GET  /v1/archive/regions/{region_id}?limit=52
GET  /v1/archive/stats
GET  /v1/grades
GET  /v1/grades/metrics
GET  /health
```

## Appendix B: FamineHypothesis Schema (Abbreviated)

```
{
```

```
    "hypothesis_id": "CERES-HYP-SDN-20260303-A7F3C1",
    "region_id": "SDN",
    "region_name": "Sudan",
    "reference_date": "2026-03-03",
    "forecast_horizon_days": 90,
    "alert_tier": "TIER-1",
    "composite_stress_score": 0.847,
    "famine_probability": {
      "p_ipc3plus_90d": 0.966,
      "p_ipc4plus_90d": 0.724,
      "p_famine_90d": 0.183,
      "sensitivity_interval_low": 0.891,
      "sensitivity_interval_high": 0.991,
      "interval_type": "input_perturbation_90pct",
      "ci_method": "expert_prior_logistic_perturbation_v2"
    },
    "driver_clusters": [
      { "driver_type": "CONFLICT",    "intensity": 0.94, "confidence": 0.91 },
      { "driver_type": "IPC_TREND",   "intensity": 0.88, "confidence": 0.95 },
      { "driver_type": "FOOD_ACCESS", "intensity": 0.76, "confidence": 0.82 }
    ],
    "ipc_phase_forecast": 4,
    "created_at": "2026-03-03T06:12:44Z"
}
```

Note: the **ci_method** field was previously named **method** in earlier API versions. The current field name is *ci_method*; the value *expert_prior_logistic_perturbation_v2* is retained for continuity and reflects the legacy characterisation — see Section 3.3 for the current description of these coefficients as author-specified initial values.